\documentclass[doublecol]{epl2}
\usepackage{amsmath}
\usepackage{amssymb}
\usepackage{epsfig}
\usepackage{amscd}
\usepackage{color}
\usepackage{comment}
\usepackage{ulem}
\usepackage{bm}
\usepackage{pdfsync}

\newcommand{\elabel}[1]{\label{e:#1}}

\newcommand{\eq}[1]{Eq.~(\ref{e:#1})}

\newcommand{\ptilde}{\tilde{p}}

\newcommand{\eqq}[1]{Equation~(\ref{e:#1})}
\newcommand{\eqtwo}[2]{Eqs~(\ref{e:#1}) and~(\ref{e:#2})}
\newcommand{\fig}[1]{Fig.~\ref{f:#1}}

\newcommand{\quot}[1]{``#1''}

\newcommand{\SET}[1]{\{#1\}}

\newcommand{\VEC}[1]{\mathbf{#1}}

\newcommand{\xvec}{\VEC{x}}

\newcommand{\deltavec}{\boldsymbol{\delta}}

\newcommand{\expb}[1]{\exp \glb #1 \grb} 
\newcommand{\glb}{\left(}  
\newcommand{\grb}{\right)}  

\newcommand{\TO}{,\ldots,}

\newcommand{\mean}[1]{\left\langle #1 \right\rangle}

\newcommand{\taucoup}{\tau_{\text{coup}}}  
\newcommand{\taucorr}{\tau_{\text{corr}}}  

\newcommand{\bds}{\beta_\text{ds}}
\newcommand{\eds}{\eta_\text{ds}}

\newcommand{\wfigureh}[3][\columnwidth]{
\begin{figure}[h]
  \onefigure[width=#1]{Figures/#2.eps}
  \caption{#3}
  \label{f:#2}
\end{figure}}

\title{Damage spreading and coupling in Markov chains}

\author{Etienne P. Bernard\thanks{E-mail:
    \email{etienne.bernard@ens.fr}}, C\'edric Chanal and Werner
  Krauth\thanks{E-mail: \email{werner.krauth@ens.fr}}}

\institute{Laboratoire de Physique Statistique, CNRS, UPMC, Ecole Normale
Sup\'{e}rieure, 24 rue Lhomond, 75231 Paris Cedex 05, France}


\date{\today}
\abstract{
  In this paper, we relate the coupling of Markov chains, at the basis
  of perfect sampling methods, with damage spreading, which 
  captures the chaotic nature of stochastic dynamics. For
  two-dimensional spin glasses and hard spheres we point out that the
  obstacle to the application of perfect-sampling schemes is posed by
  damage spreading rather than by the survey problem of the entire
  configuration space. We find dynamical damage-spreading
  transitions deeply inside the paramagnetic and liquid phases, and we
  show that critical values of the transition temperatures and
  densities depend on the coupling scheme. We discuss our findings in
  the light of a classic proof that for arbitrary Monte Carlo
  algorithms damage spreading can be avoided through
  non-Markovian coupling schemes.}

\begin{document}
\maketitle
\section{Introduction}

Chaos manifests itself in Hamiltonian dynamical systems when any two
nearby initial configurations drift apart with time. Chaos can also
be defined for cellular automata and for Markov chain algorithms. In
these dynamical systems, following Kauffman \cite{Kauffman_1969},
the drifting-apart of configurations is termed \quot{damage spreading}.
In contrast, for \quot{regular} dynamics, two nearby initial configurations
become identical after a finite time, and remain indistinguishable from
then on. For Markov-chain Monte Carlo algorithms, the closely related
case where the entire space of initial configurations becomes identical is
termed \quot{coupling}. Once it has coupled, the Markov chain has lost all
correlations with the initial configuration. The coupling of Markov chains
has risen to great prominence when Propp and Wilson used it for a perfect
sampling method for Markov chains named \quot{Coupling From The Past}
(CFTP)\cite{Propp_Wilson_1996}. When applicable, this method overcomes the
problem of estimating the correlation time of a Monte Carlo calculation.
In the present article, we shall discuss the fruitful connection
between damage spreading and coupling \cite{Chanal_Krauth_2008}.

In systems with $N$ elements (spins, hard spheres, etc), the
configuration space generally grows exponentially with $N$, and CFTP
thus faces two distinct challenges. First, it must \textit{survey}
the entire configuration space in order to prove coupling. Second,
it must avoid damage spreading which would cause the coupling time to
\textit{explode}: it would become much larger than the correlation time
as any two configurations have a very small probability for finding each
other in a large space.

The surveying problem is avoided in systems with a special property
called \quot{partial order}, as for example the ferromagnetic Ising model
under heat-bath dynamics\cite{Propp_Wilson_1996,SMAC}. For more general
systems (without partial order, but with local update algorithms), such as
spin glasses and hard spheres, a recent \quot{patch} algorithm inspired by
numerical block scaling ideas \cite{Chanal_Krauth_2008,Chanal_Krauth_2010}
allows us to rigorously follow a superset of all initial conditions until
it couples. This algorithm generates only modest overheads of memory and
CPU time \cite{Chanal_Krauth_2008}. It was found that the coupling can be
established after a time evolution very close to the coupling time.

The second problem, the explosion of the coupling time related to damage
spreading, poses the veritable obstacle to the application of CFTP
ideas. Damage spreading has been studied in many physical systems, in
particular spin glasses \cite{Derrida_Weisbuch_1987}. In
several spin glass models with heat-bath dynamics, it is now well
established that a dynamical damage-spreading transition occurs at a
critical temperature, $\bds$, located in the paramagnetic
phase\cite{Campbell_1991}: the dynamic is regular at temperatures
higher than $1/\bds$ and chaotic at lower temperatures. Even for the
two-dimensional $\pm J$ Ising spin glass, which has a thermodynamic
phase transition at $T=0$ \cite{Morgenstern_1979, McMillan_1983,
  Singh_1986, Bhatt_Young_1988}, the transition to chaos takes place
at finite temperature \cite{Campbell_1991}. We study in this paper the
damage spreading of spin glasses and hard spheres, the divergence of the ratio of 
the coupling time and the correlation time, for different algorithms. 

\section{Random walks in high dimensions}
Before analyzing two-dimensional spin glasses and hard spheres, we
illustrate coupling and damage spreading in a simple Markov
chain algorithm that can be interpreted either as a random walk in an
$N$-dimensional hypercubic lattice, as the dynamics of $N$ distinguishable
non-interacting particles in a one-dimensional lattice of length $L$,
or as $N$ non-interacting Potts spins with $L$ states. For the random
walk (see \fig{couplage}), each $N$-dimensional lattice site
$i=\SET{i_0\TO i_{N-1}}$ is described by integers $i_k \in \SET{0\TO
L-1}$ with periodic boundary conditions in the $i_k$ . The particle
can hop from site $i$ to one of $i$'s nearest neighbors in direction $k$,
$j=i \pm \delta_k$, with $\delta_k=\SET{0,\dots,1,0,\dots}$ (periodic
boundary conditions are again understood).  The probability for moving
from $i$ to $j$ is
\begin{equation}
p(i \to j) = 
\begin{cases}
\tfrac{1}{3N} & \text{for $j=i\pm \delta_k$} \\
\tfrac{1}{3} & \text{for  $j=i$} \\
0  &\text{otherwise}
\end{cases}.
\elabel{random_walk}
\end{equation}
\wfigureh[8cm]{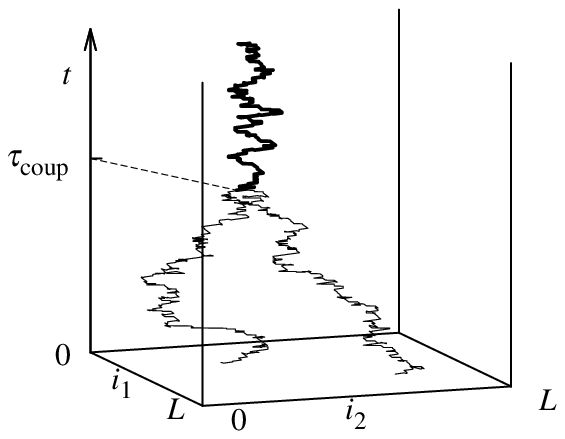}{Coupling of two random walks in a periodic
$N$-dimensional hypercubic lattice of length $L$. After the time
$\taucoup$, the two random walks evolve identically. The chaotic coupling
of \eq{coupling_random_walk_chaotic} is shown. For the regular coupling
of \eq{coupling_random_walk_regular}, the displacement at time $t$
is in the same direction $k$, and it is a function of $i_k$ only. }
The simulation thus samples at each time step one dimension, $k$,
among the $N$ available ones (it moves in \quot{$x$}, or \quot{$y$} or
\quot{$z$}, etc).  In dimension $k$, it then hops with probabilities
$1/3$ each to the left or to the right, or remains on the same
site. \eqq{random_walk} also describes $N$ distinguishable non-interacting
particles on a one-dimensional lattice of length $L$, again with periodic
boundary conditions: At time $t$, a randomly chosen particle $k$ hops
to the left or to the right, or it remains on the same site, each with
probability $1/3$, as above.

A two-configuration coupling is a random process $\ptilde(i
\to j, i'\to j')$ for the joint evolution of two random walks such
that integrating over one of them yields the original random walk of
\eq{random_walk} for the other. After they meet, the two configurations
evolve in the same way. The simplest choice for a coupling is the product ansatz,
\begin{equation}
\ptilde(i \to j, i'\to j') = 
\begin{cases}
p(i \to j) p(i' \to j') &\text{if $i \ne i'$}\\
p(i \to j)              &\text{if $i = i'$, $j=j'$}\\
0                       & \text{otherwise} 
\end{cases},
\elabel{coupling_random_walk_chaotic}
\end{equation}
where the two random walks evolve independently from each other if
they are on different sites $i$ and $i'$, but stay together once
they have met ($j=j'$ if $i = i'$). To implement this coupling for any
number of configurations, one samples at each time step independent
random moves at each site, so that particles on the same site
experience the same randomness. In the above-mentioned representation
of particles on the one-dimensional line, we consider the coupling of
two $N$-particle systems, again described by
\eq{coupling_random_walk_chaotic}, as the independent evolution, at
time $t$, of the $L^N$ possible configurations of the
system. Naturally, the coupling time scales as $L^N$ whereas the correlation time (in sweeps)
behaves as $L^2$.

\wfigureh[8cm]{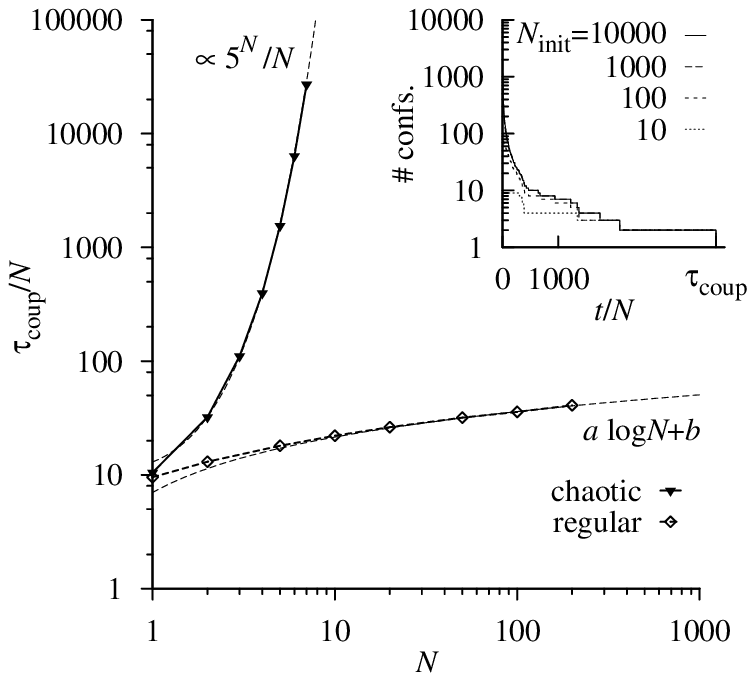}{Chaotic and regular couplings for
the random walk in a $N$-dimensional hypercubic lattice of length $L=5$
(see \eqtwo{coupling_random_walk_chaotic}{coupling_random_walk_regular},
respectively). The random process for a single random walk is defined by
\eq{random_walk} in both cases. \textit{Inset}: Number of configurations
\textit{vs.} time as a function of the number $N_\text{init}$ of initial
configurations, for $N=6$, $L=5$ for the chaotic coupling. The same
realization of the coupling of \eq{coupling_random_walk_regular} is used
for all runs.}

An alternative coupling consists in sampling, at time $t$, one dimension
$k$ common to all random walks. The two-configurations coupling scheme is then
\begin{multline}
\ptilde(i_k \to j_k, i'_{k}\to j'_{k}) \\= 
\begin{cases}
p(i_k \to j_k) p(i'_{k} \to j'_{k}) &\text{if $i_k \ne i'_{k}$}\\
p(i_k \to j_k)                      &\text{if $i_k = i'_{k}, j_k = j'_{k} $}\\
0                       &\text{otherwise}
\end{cases}
\elabel{coupling_random_walk_regular}
\end{multline}
so that two configurations $i$ and $j$ with $i_k = i'_k$ will preserve
this common coordinate ($j_k = j'_{k}$). In the representation of
$N$ particles on a one-dimensional lattice, the same particle $k$ is
selected for each configuration, and for two different configurations,
the particles labelled $k$ stay together once they have met on the same
site. The dynamics is then regular and the coupling time is
$\taucoup/N \sim a \log{N}$ (see \fig{rw_coupling_time}). The
logarithmic behaviour is explained by the fact that particles move
independently from each other, the coupling time for the entire
system is thus the maximum of the $N$ coupling times for each particle.

In conclusion, we see that the same $N$-dimensional random walk of
\eq{random_walk}, with a correlation time of order $L^2$, allows two
very different coupling, chaotic and regular. In spin glasses and hard
spheres, these regimes are realized for the same coupling at different
temperatures.

\section{Spin glass}

The random walk considered previously can also be considered
as an $L$-state Potts model at infinite temperature evolving under
heat-bath dynamics.

The product ansatz of \eq{coupling_random_walk_chaotic} would correspond
to the independent evolution of the spin configurations, and it is
clearly chaotic. With the coupling of 
\eq{coupling_random_walk_regular}, in contrast, all spins evolve and
couple independently at $\beta=0$, and the global
coupling time is again the maximum of the coupling times of the
individual spins. The Monte Carlo dynamics is thus regular,
and the diagram of \fig{rw_coupling_time} carries over to the general
case with $L \ge 2$. At finite temperatures $\beta$, the energy of a
spin configuration is given by
\begin{equation*}
   E= - \sum_{\langle i,j\rangle} J_{ij} s_i s_j.
\end{equation*}
We first consider heat-bath dynamics, which consists in choosing one
spin $s_k$ and updating it with probabilities
\begin{equation}
  \pi(s_k = \pm 1) =  \frac{1}{1 + \expb{ \mp 2 h_k \beta} },
  \elabel{}
\end{equation}
where  the field on site $k$ is given by $h_k = \sum_{l} J_{kl} s_l$. The
coupling is defined by the use of the same random numbers for each
configurations.

For the two-dimensional ferromagnetic Ising model (all $J_{ij}=1,
L=2$), the dynamics remains regular at all temperatures. Below the
Curie temperature, $\taucoup$  is very large, but so is the correlation
time $\taucorr$, and the partial order implies that the complexity
of $\taucoup/\taucorr$ $\leq O(\log{N})$ \cite{Propp_Wilson_1996}.
The partial order is preserved in the disordered Ising model
with ferromagnetic interactions $J_{ij}=J_{ji} \ge 0$, and in this model
also, the theorem of Propp and Wilson guarantees that $\taucoup $ is,
up to logarithms, of the same order as $\taucorr$.

Frustrated models, as for example spin glasses, do not
exhibit partial order, and can thus undergo a damage-spreading
transition. In the two-dimensional $\pm J$ Ising spin glass, the
quenched random interactions satisfy $J_{ij}=J_{ji} = \pm1$ with
equal probability. Although this model is paramagnetic for all finite
temperatures, Campbell and de Arcangelis\cite{Campbell_1991} found a
damage-spreading transition for the heat-bath algorithm at $\bds \simeq
0.59$. In previous work \cite{Chanal_Krauth_2008,Chanal_Krauth_2010},
we succeeded in coupling large systems down to this temperature using
the patch algorithm. We showed that the patch algorithm's upper bound
on the coupling times agrees well with the lower bound obtained from
a partial-coupling approach, where one checks coupling for a finite
number $N_\text{init}$ of random initial conditions rather than for the
entire configuration space ($N_\text{init}=2^N$). As shown in the inset
of \fig{Spin_glass_coupling_time_N}, for one realization of the random
process, the coupling time does not vary if $N_\text{init} \gtrsim 10$,
and for $N_{\text{init}}= 1000$ it equals the coupling time for the
entire configuration space.

\wfigureh[8.8cm]{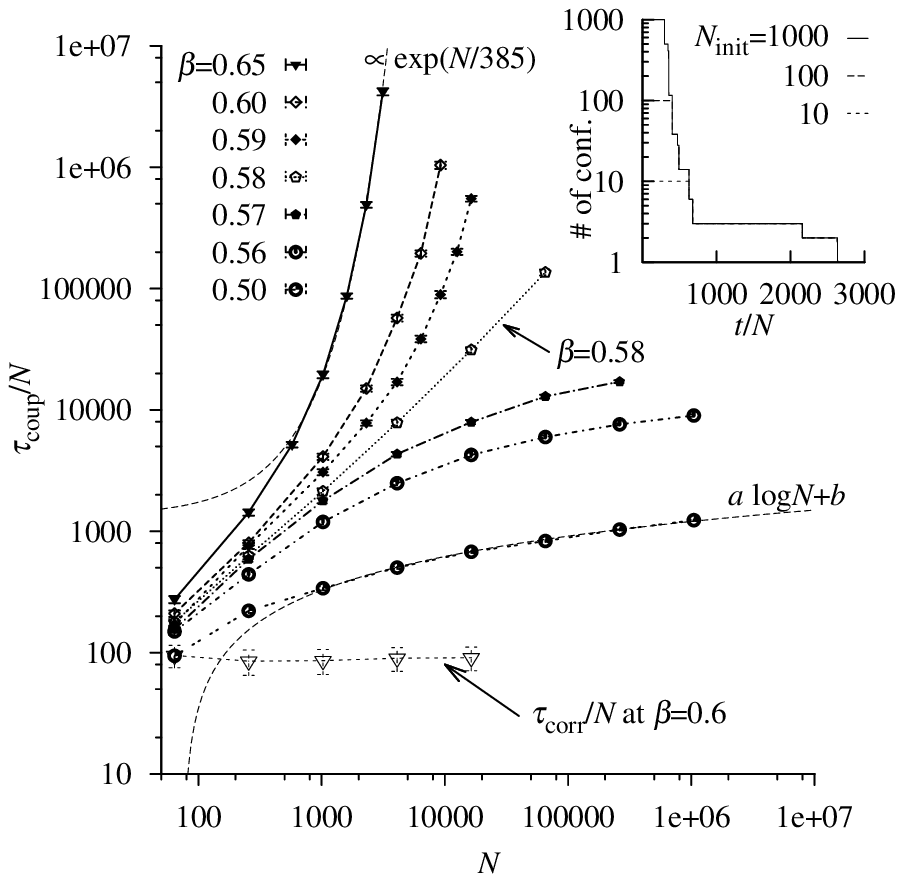}{Disorder-averaged coupling
time for the heat-bath algorithm of the two-dimensional $\pm J$ Ising
spin glass. A dynamical phase transition is seen at the damage-spreading
temperature $\bds \simeq 0.58$. Inset: Saturation phenomenon for $N=64^2$ spins at $\beta=0.56$.}

In the main graph of \fig{Spin_glass_coupling_time_N}, we show the coupling
time as a function of $N$ at constant temperature. A
dynamical phase transition is seen at the damage-spreading
temperature $\bds \simeq 0.58$. In the chaotic phase, $\taucoup/N$ grows exponentially with $N$,
but only logarithmically in the regular phase. 
The dynamical phase transition in this model (without a spin glass phase
at finite $\beta$), although not mathematically proven, appears firmly
established. It is without influence on single-particle properties. To
illustrate this point, we verify that the correlation time
$\taucorr/N$, computed with the autocorrelation function 
\begin{equation}
q(t)=\frac{1}{N} \sum_{i=0}^N  \mean{s_i(0)s_i(t)},
\elabel{}
\end{equation}
remains constant in the chaotic phase and only $\taucoup/\taucorr$
diverges with $N \to \infty$. 

After the heat-bath algorithm, we now discuss the Metropolis algorithm,
where individual spins $s_k$ are flipped with a probability depending
on their local field. In the standard implementation, spin flips are
accepted with a probability equal to $1$ at infinite temperature. To
allow coupling at any $\beta$ we use
\begin{equation}
p(s_k \to -s_k) = \frac{2}{3} \min(1,\exp(-2 \beta s_k h)).
\elabel{}
\end{equation}
At each step the same spin $k$ is updated for all copies of the
system. For this dynamics, several coupling schemes can be set up. If the same
random number $\gamma$ is used for each 
configuration, the coupling does not take place, as two opposite
configurations will always stay opposite. We adapt the regular
coupling of \eq{coupling_random_walk_regular} and use two
independent random numbers, $\gamma_1$ for \quot{up} spins and $\gamma_2$ for
\quot{down} spins. The coupling time has then the same qualitative
behaviour as in \fig{Spin_glass_coupling_time_N}, logarithmic at high temperatures and
exponential at low temperatures, but with a critical temperature
$\bds \simeq 0.33$. The Metropolis algorithm, with this coupling scheme,
has thus a higher dynamical critical temperature than the heat-bath
algorithm, with which it shares all the qualitative features. This
confirms that the dynamic damage-spreading transition is algorithm
dependent. One may also choose the random numbers in the Metropolis algorithm
using $\gamma$ for $s_k=1$ and $1-\gamma$ for
$s_k=-1$. This scheme correlates opposite
spins better and the critical temperature is found to be $\bds \simeq 0.52$. This
results shows that, like for the previous random walk,
the same Markov-chain allows for qualitatively different coupling.

\section{Hard spheres}
After spin glasses, we now consider another key model in statistical
physics, namely hard spheres. This model's Hamiltonian dynamics, realized in the event-driven
molecular dynamics algorithm \cite{Alder_1957,SMAC}, is chaotic
for all densities  and for all $N$ \cite{Sinai,Simanyi,SMAC}.  In this
section, we will present several Monte Carlo algorithms for hard
spheres, which allow for coupling of the entire configuration space.
Two of the algorithms remain regular below a finite critical packing
fraction, $\eds$, in the limit $N \to \infty$. In the following discussion we are
not concerned with algorithmic efficiency of the implementation, and only
concentrate on the coupling properties.

\subsection{Birth-and-death algorithm}

In the grand-canonical birth-and-death Monte Carlo algorithm, particles
are placed  inside the box at random positions $\xvec=(x_k, y_k)$ at rate
$\lambda$ if no overlaps with previously placed disks are generated.
The life time of each disk is sampled from an exponential distribution
with rate $1$. One realization of the algorithm is represented in the
diagram of \fig{Birth_and_death}.  The mean number $\mean{N}$ of particles
in this system is controlled by the activity $\lambda$.

\wfigureh[8cm]{Birth_and_death}{Grand-canonical birth-and-death algorithm
   for hard disks. Disk $i$ appears at time $t_i$, at
   position $\xvec_i=(x_i,y_i)$, and it disappears at time $t_i +
   \tau_i$. In time, disk $i$ describes a cylinder. Disks (cylinders) which
   are accepted, because they create no overlaps with earlier disks,
   are drawn in dark gray, while rejected disks are drawn in light gray.
   The configuration space of this system is infinite, yet the possible
   configurations at time $t$ are a subset of the finite set (of dark
   and light cylinders) produced from a horizontal cut in this diagram.}

This model's state space is infinite, but the survey of
all possible initial conditions is nevertheless feasible
\cite{Wilson_2000,Chanal_Krauth_2010}. For any realization of the
algorithm, the possible configurations at time $t$ are a subset
of the finite set produced from a horizontal cut in the diagram of
\fig{Birth_and_death}. The patch algorithm again yields sharp upper
bounds for $\taucoup$ \cite{Chanal_Krauth_2010}. Surprisingly, this
algorithm for hard disks remains regular below a finite density $\eds$
in the limit $N\to \infty$ \cite{Wilson_2000,Kendall_2000}.

\wfigureh[8cm]{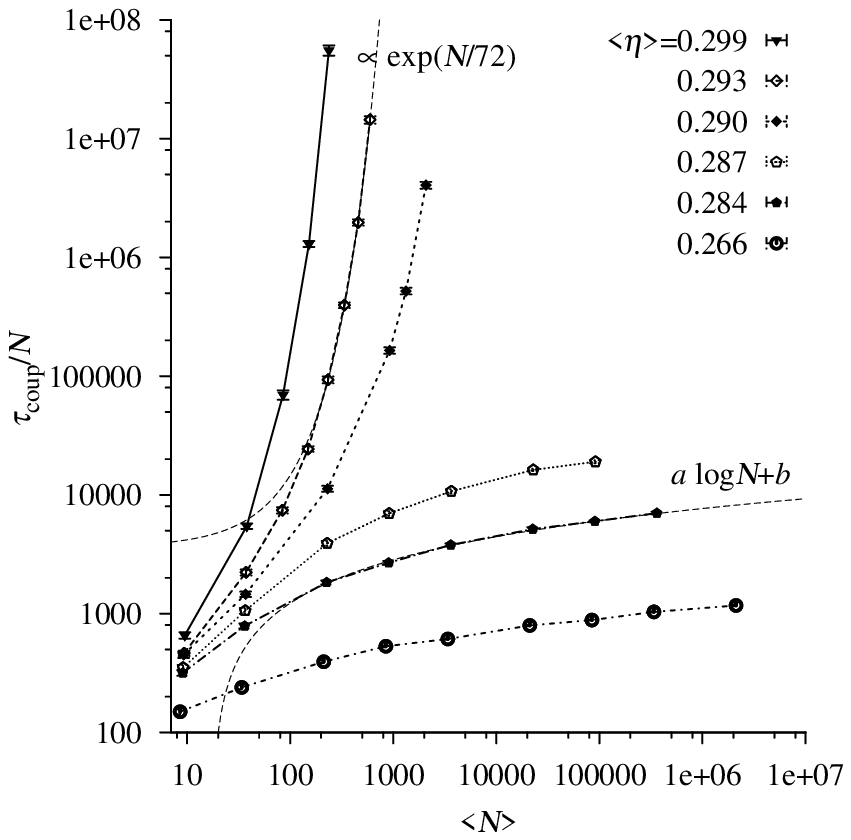}{Coupling time of the birth-and-death
   algorithm of \fig{Birth_and_death} for two-dimensional hard spheres
   (estimated with  $N_\text{init} = 100$). The damage spreading
   transition occurs at a packing fraction $\eds \simeq 0.29$.}

We again study damage spreading in this model by applying the same
Monte Carlo dynamics (same choice of $\xvec_i, t_i, \tau_i$) to
$N_\text{init}$ random hard-sphere initial conditions at time $t=0$
with life times sampled from an exponential distribution.  The data shown
in \fig{BD_coupling_time_N} again indicate a dynamical phase transition
between the regular regime at packing fractions $\eta < \eds \simeq 0.29$
and the chaotic regime above $\eds$. This density corresponds to the
limiting density found with the patch algorithm \cite{Chanal_Krauth_2010}.

\subsection{"Labelled displacement" algorithm}

A canonical version of the birth-and-death algorithm is the
\quot{labelled displacement} algorithm where, at times $t = 0,1,2,
\dots$, a randomly chosen particle $k$ is moved to a
random position $\xvec_k$, if this move creates no overlaps.
\wfigureh[8cm]{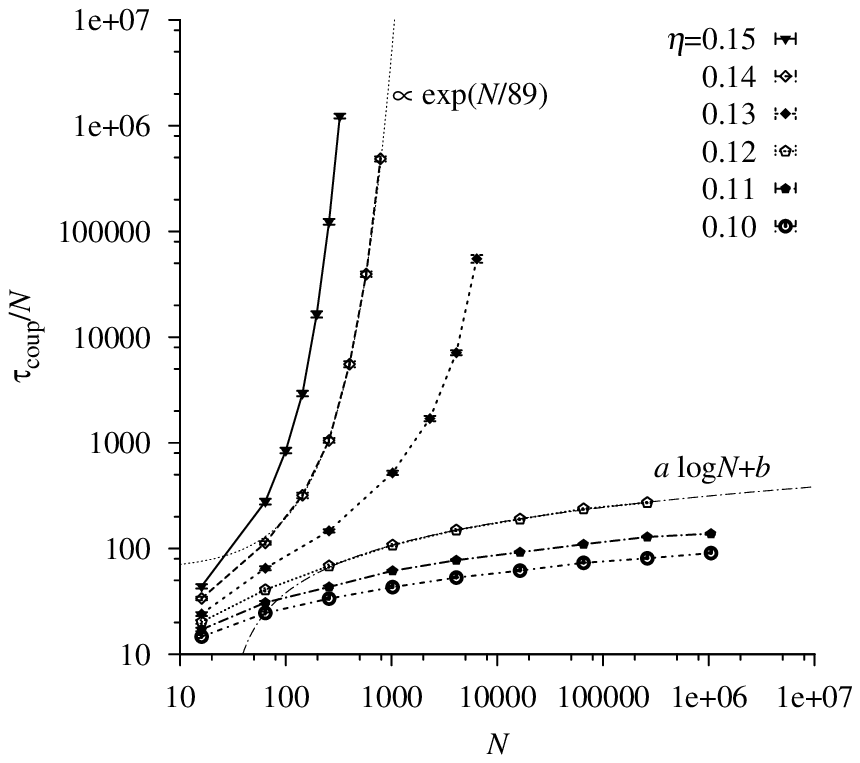}{Coupling time for the labelled
   displacement algorithm. The dynamical transition to chaos occurs
   at a lower density ($\eds \simeq 0.13$) than for the birth-and-death algorithm of
   \fig{BD_coupling_time_N}.}
We see clear evidence of a dynamical phase transition at a critical density $\eds
\simeq 0.13$ (see \fig{label_coupling_time_N}), which is smaller than for the closely related
birth-and-death algorithm.

\subsection{Spot algorithm} 

We finally study the coupling for a Markov-chain similar to the
Metropolis algorithm: the spot algorithm. 
\wfigureh[7cm]{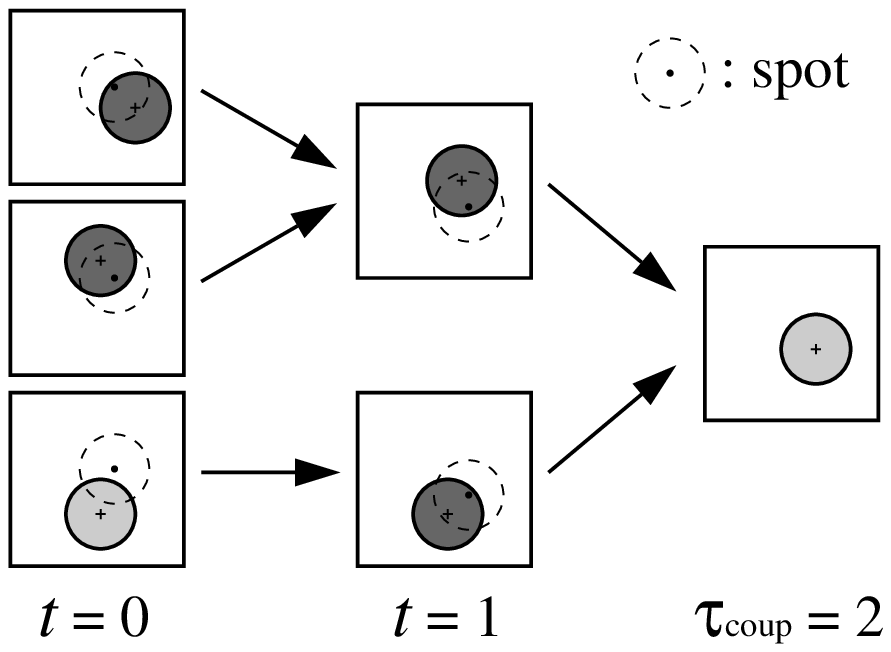}{Spot algorithm for hard spheres:
   The randomly chosen spot position defines the attempted move of a
   disk inside the spot. The
   spot radius satisfies $\sigma_\text{spot} \le
   \sigma$, and at most one disk is moved at time $t$. An
   example with $N=1$ and $\sigma_\text{spot}= \sigma$ is shown.}
The Metropolis algorithm for
$N$ hard spheres consists in moving, at time $t$, a particle $k$  by a
random vector $\deltavec=(\delta_x,
\delta_y)$. As the configuration space is continuous, the coupling
probability is zero if one uses a naive coupling scheme. The following
spot algorithm is more successful (although we will show its coupling
is chaotic at all densities): at time $t$, it places a spot,
a disk-shaped region with radius $\sigma_{\text{spot}} \le \sigma$, at a
random position $\xvec_s$. 
\wfigureh[8.5cm]{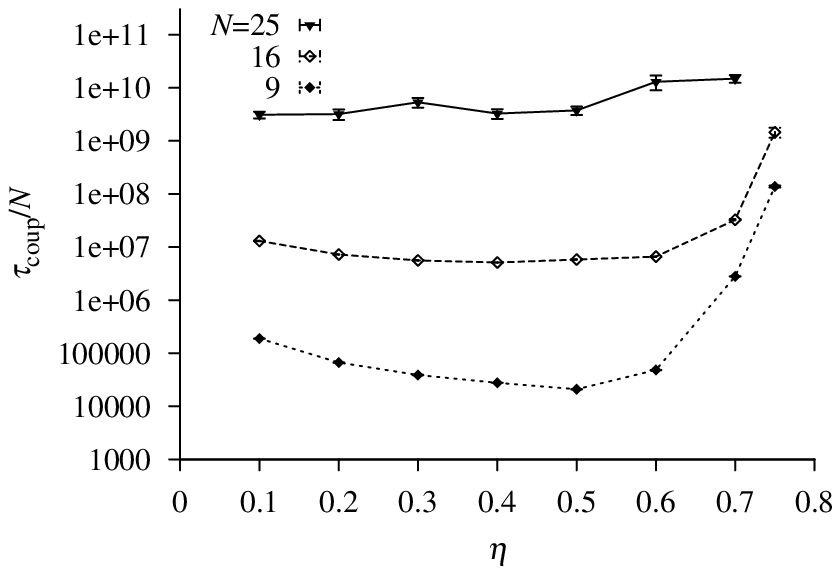}{Coupling time $\taucorr$  of
   the spot algorithm for $N$ hard disks (estimated with $N_{\text{init}}=
   100$) as a function of packing fraction $\eta$. For all $\eta$, the
   coupling time is exponential in $N$, and we conjecture the coupling to
   be chaotic.}
The spot contains at most one disk center,
and the move consists in placing this disk at $\xvec_s$, if this creates
no overlap with other particles (See \fig{spot_moves}). The spot
algorithm satisfies detailed balance, and it generates the same moves
as the Metropolis algorithm. Moreover, as illustrated in
\fig{spot_moves}, it succeeds in coupling.
However, as shown in \fig{Spot_coupling_time_eta}, the coupling time of
the spot algorithm is an exponential function of $N$ for all densities:
the coupling is always chaotic.

\section{Conclusion}

In conclusion, we have in this paper studied the relationship between
the coupling of Markov chains, which is of critical importance for the
subject of perfect sampling, and damage spreading, which
exposes the chaotic nature of the Monte Carlo dynamics. 

For the two-dimensional $\pm J$ Ising spin glass, which lacks an equilibrium phase
transition at finite temperatures, we confirm the existence of a
dynamical phase transition at $\bds \simeq 0.58$
\cite{Campbell_1991} for the heat-bath algorithm. For lower
temperatures the coupling time explodes. The Metropolis algorithm has
the same damage spreading behaviour but with higher critical temperatures:
$\bds \simeq 0.33$ or $\bds \simeq 0.52$ for two simple
coupling schemes. All damage-spreading transitions for this system are
deeply inside the paramagnetic phase.

For the two-dimensional hard-sphere system, we analyzed three local
Monte Carlo algorithms, the birth-and-death
algorithm, inspired from Poisson point processes, its canonical
version (the \quot{labelled displacement} algorithm), and the spot algorithm,
a straightforward adaptation of the Metropolis algorithm. The first
algorithm shows a regular regime only for packing densities below  $\eds \simeq
0.29$, the coupling time was then of the same order of
magnitude as the correlation time. The canonical version of the
birth-and-death algorithm had a critical density of $\eds \simeq
0.13$. These transition densities are again deeply in the liquid phase.

Both for spin glasses and for hard spheres, the
rigorous survey of the configuration space\cite{Chanal_Krauth_2008} remains
feasible for all temperatures and densities. The application of
perfect sampling methods to these challenging problems is thus
not so much limited by the surveying problem, as the patch algorithm
allows to track the evolution of the entire configuration space, but
more by damage spreading, the underlying chaotic nature of the Monte Carlo dynamics.

In this context, it is of great interest that Griffeath
\cite{Griffeath_1975} has constructed a coupling that always remains
regular: It realizes the coupling at time $t$ and at position $X_t$
of two Markov chains that have started at time $t=0$ at configurations
$X_0$ and $X_0'$ with the minimum of the probabilities to go from $X_0
$ or from $X'_t$ to $X_t$. Griffeath's coupling is non-Markovian and
very difficult to construct in practice, but it may point the way to
couplings that remain regular at lower temperatures and higher densities
than the naive couplings we discussed in this paper.

\acknowledgments
We thank Alistair Sinclair for helpful correspondence.

\bibliography{biblio}
\bibliographystyle{eplbib}

\end{document}